\documentclass[aps,onecolumn,showpacs,preprint]{revtex4}
\usepackage{amssymb}
\usepackage{amsmath}
\usepackage{graphicx}

\newcommand{\be}{\begin{eqnarray}}
\newcommand{\ee}{\end{eqnarray}}
\newcommand{\ca}{\mathbb{C}^\dag}
\newcommand{\cb}{\mathbb{C}}

\begin{document}
\title[]{Quantum Spin Hall Effect in Graphene Nanoribbons: Effect of Edge Geometry}

\author{Jun-Won \surname{Rhim}}
\affiliation{Department of Physics and IPAP, Yonsei University,
Seoul, 120-749, Korea }
\author{Kyungsun \surname{Moon}}
\affiliation{Department of Physics and IPAP, Yonsei University,
Seoul, 120-749, Korea }

\begin{abstract}
There has been tremendous recent progress in realizing topological
insulator initiated by the proposal of Kane and Mele for the
graphene system. They have suggested that the odd $Z_2$ index for
the graphene manifests the spin filtered edge states for the
graphene nanoribbons, which lead to the quantum spin Hall
effect(QSHE). Here we investigate the role of the spin-orbit
interaction both for the zigzag and armchair nanoribbons with
special care in the edge geometry. For the pristine zigzag
nanoribbons, we have shown that one of the $\sigma$ edge bands
located near $E=0$ lifts up the energy of the spin filtered chiral
edge states at the zone boundary by warping the $\pi$-edge bands,
and hence the QSHE does not occur.
Upon increasing the carrier density above a certain critical
value, the spin filtered edge states are formed leading to the
QSHE. We suggest that the hydrogen passivation on the edge can
recover the original feature of the QSHE. For the armchair
nanoribbon, the QSHE is shown to be stable. We have also derived
the real space effective hamiltonian, which demonstrates that the
on-site energy and the effective spin orbit coupling strength are
strongly enhanced near the ribbon edges. We have shown that the
steep rise of the confinement potential thus obtained is
responsible for the warping of the $\pi$-edge bands.

\end{abstract}



\maketitle

\section{Introduction}
Recently, the role of intrinsic spin-orbit coupling(SOC) in
graphene and graphene multilayer systems has attracted a lot of
attention as one of the model system to realize the new quantum
state of matter and also for the possible spintronics
application\cite{a1,a2,a3,a4,a5,a6,a7,a8,a9,a10,a11,a12,a13,a14,a15,a16,a17,a18,a19,a20,a21,a22,a23,a24,a25,a26}.
It has been well known that upon including the SOC, graphene
supports the quantum spin Hall effect(QSHE). This implies that
there exist spin filtered chiral states at the edges of graphene
leading to the nonzero quantized spin current\cite{a1,a2,a3}. It
has been argued that the odd $Z_2$ index of the bulk graphene
guarantees the existence of the QSHE\cite{a1,a10,a16,a27,a28}.
This integral number represents the inherent topological
properties of graphene just as the Chern number characterizes the
quantum Hall systems\cite{a29,a30}. Furthermore this interesting
role of the SOC in graphene has stimulated the studies on the
novel kind of material called the topological
insulator(TI)\cite{t1,t2,t3,t4}. Following the initial theoretical
proposal to graphene, several other materials have been
theoretically suggested to be possible candidates of TI, which
have been experimentally confirmed later on\cite{t2,t3}. In
contrast, the QSHE of graphene has not been experimentally
observed yet. This may be due to the weak intrinsic SOC in
graphene so that the energy dispersion of the edge state is too
small to be resolved at experimentally feasible temperature
scale\cite{a5,a6,a7}. More interestingly, one can speculate that
the graphene nanoribbon system may demonstrate the unexpected
interplay between the topological properties based on the bulk
energy band and the edge geometry of the nanoribbon. In this
regard, the graphene nanoribon system will serve as a standard
model whose edge geometry can be well characterized and controlled
by a few parameters\cite{a10}.

The existence of the QSHE in graphene as mentioned above is mostly
based on the study of the Kane-Mele(KM) model which has received
considerable attention as a realization of the Haldane's original
idea about the quantum Hall effect without magnetic field in the
honeycomb lattice\cite{haldane}. In the KM-hamiltonian, the low
energy processes between $p_z$ orbitals mediated by the SOC are
described by the imaginary next nearest neighbor(n.n.n.) hopping
terms
$i\xi_1/3\sqrt{3}\nu_{ij}s^z_{\alpha\beta}c^\dag_{i\alpha}c_{j\beta}$\cite{a1,a2}.
Based on the group theoretical and perturbative arguments, it has
been elucidated that the SOC term for the nearest neighbor(n.n.)
hopping process vanishes at the Dirac point by the lattice
symmetry of graphene and hence the leading SOC term originates
from the n.n.n. hopping processes\cite{a5,a6,nnn1}. Several
authors have indicated that one can obtain much more enhanced
intrinsic spin orbit coupling(ISOC) such as the new hopping
processes in the bilayer graphene system, which arise due to the
different lattice symmetries around each carbon atom\cite{b1,b2}.
Here we want to emphasize that in the graphene nanoribbons, one
should be very careful in applying the low-energy effective
hamiltonian to the edges of the graphene, where the bulk lattice
symmetry is broken.

In the paper, we investigate the effect of the edge geometry on
the low energy physics of the intrinsic SOC both in the zigzag and
armchair nanoribbons based on the tight binding hamiltonian. In
order to describe the SOC, we have included the $s, p_x, p_y$ and
$p_z$ orbitals of carbon atoms instead of using the effective
Kane-Mele term containing the $p_z$ orbital alone. For the
pristine zigzag nanoribbon, we have demonstrated that one of the
$\sigma$ edge bands made of the $s, p_x, p_y$ orbitals located
near $E=0$ lifts up the energy of the spin filtered chiral edge
states at the zone boundary, and hence the QSHE does not occur. By
increasing the carrier density within a certain range, the system
exhibits the QSHE with spin filtered chiral edge states. By
further increasing the carrier density above a certain critical
value, the QSHE disappears again by adding an extra pair of edge
states leading to the even number of edge states at each side of
the edges. We have also studied the role of hydrogen passivation
on the edges, whose orbitals hybridize with the $\sigma$ edge
bands located near $E=0$ and then the two edge bands are repelled
from each other by creating large energy gaps. Remarkably, we have
noticed that the original feature of the QSHE revives with
hydrogen passivation. For the armchair graphene nanoribbon(AGNR),
we have noticed that the QSHE is mostly quite stable with or
without passivation. However the edge state of the armchair
nanoribbon is too widely spread from the edge on the order of
$(\gamma_0/\xi_1)a\cong 1mm$, where $\gamma_0$ represents the
nearest neighbor hopping amplitude of the $\pi$ electrons, $\xi_1$
the SOC induced next nearest neighbor hopping amplitude, and $a$
the lattice spacing.

In the section II, we have introduced the tight binding
hamiltonian, which describes the graphene nanorribbon system. The
hamiltonian includes both the $p_z$ and the $s, p_x, p_y$ orbitals
of carbon atom, the atomic spin-orbit coupling term, and the edge
passivation term. In the section III, we have investigated the
band structure of the pristine ZGNR and then the band structure of
the AGNR is subsequently calculated. In the section IV,  the
effect of hydrogen passivation on the band structure of the ZGNR
is studied. In the section V, we have constructed the real space
effective hamiltonian for the ZGNR. The summary will follow in the
section VI.

\section{The tight binding hamiltonian}
The $\pi$ and $\sigma$ bands of the ZGNR can be obtained by the
following tight binding hamiltonian \be H=H_\pi+H_\sigma+H_{SO}
\ee where $H_\pi$ represents the nearest neighbor hopping
processes between $p_z$ orbitals, $H_\sigma$ the matrix elements
among $s$, $p_x$, and $p_y$ orbitals, $H_{SO}$ the on-site atomic
spin orbit coupling term which connects the two Hilbert spaces
together. The hamiltonian $H_\pi$ for the $p_z$ orbitals of the
ZGNR is given as follows\cite{neto} \be
H_\pi=\gamma_0\sum_{\langle
i,j,m,n\rangle}[C^{\dagger}_{p_z,B,j,n} C_{p_z,
A,i,m}+\mathrm{h.c.}],\ee where the indices $A(B)$, $i$($j$) and
$m$($n$) represent the sublattices, dimer lines, and unit cells
along the $y$-direction respectively as shown in Fig. 1 and
$\gamma_0=V_{pp\pi}=-3.03\mathrm{eV}$. The angular bracket under
the summation stands for the nearest neighbor pairs. The
hamiltonian $H_\sigma$ can be written by \be
H_\sigma=\sum_{\alpha,i,m}\ca_{\alpha,i,m}\mathbb{E}_0\cb_{\alpha,i,m}+\sum_{\langle
i,j,m,n\rangle}[\ca_{B,j,n}\mathbf{\Sigma}_l\cb_{A,i,m}+\mathrm{h.c.}]\ee
where the index $\alpha$ represents the sublattices $A, B$. The
on-site spin-orbit coupling term is given by
$H_{SO}=\frac{\xi_0}{2}\sum\vec{L}\cdot\vec{s}$. We have omitted
the summation over the spin indices in $H_\sigma$, since the
hamiltonian is invariant over spin. $\cb_{\alpha,i,m}$ is a column
vector of a form $[c_{p_y}~ c_{p_x}~ c_{s}]^{\mathrm{T}}$ at each
site $\{\alpha,i,n\}$, while $\mathbb{E}_0$ and
$\mathbf{\Sigma}_l$ represent the on-site and hopping matrices
respectively with $l=(1+i-j)(m-n+\frac{3-(-1)^i}{2})$, which have
the following matrix elements for the case of zigzag termination
\be
\mathbb{E}_0=\left(\begin{array}{ccc}0~&0~&0\\0~&0~&0\\0~&0~&\epsilon_s\end{array}\right)\ee
\be
\mathbf{\Sigma}_0=\left(\begin{array}{ccc}V_{pp\pi}&0&0\\0&-V_{pp\sigma}&V_{sp\sigma}\\0&-V_{sp\sigma}&V_{ss\sigma}\end{array}\right)\ee
\be\mathbf{\Sigma}_1=\left(\begin{array}{ccc}(V_{pp\pi}-3V_{pp\sigma})/4&\sqrt{3}(V_{pp\pi}+V_{pp\sigma})/3&\sqrt{3}V_{sp\sigma}/2\\\sqrt{3}(V_{pp\pi}+V_{pp\sigma})/3&(3V_{pp\pi}-V_{pp\sigma})/4&-V_{sp\sigma}/2\\
-\sqrt{3}V_{sp\sigma}/2&V_{sp\sigma}/2&V_{ss\sigma}\end{array}\right)\ee
\be\mathbf{\Sigma}_2=\left(\begin{array}{ccc}(V_{pp\pi}-3V_{pp\sigma})/4&-\sqrt{3}(V_{pp\pi}+V_{pp\sigma})/3&-\sqrt{3}V_{sp\sigma}/2\\-\sqrt{3}(V_{pp\pi}+V_{pp\sigma})/3&(3V_{pp\pi}-V_{pp\sigma})/4&-V_{sp\sigma}/2\\
\sqrt{3}V_{sp\sigma}/2&V_{sp\sigma}/2&V_{ss\sigma}\end{array}\right).\ee
Here, $\epsilon_s=-8.87\mathrm{eV}$ is the on-site energy of the
$s$ orbital relative to that of the $p$ orbital and the various
hopping parameters are chosen to be $V_{pp\pi}=-3.03,
V_{pp\sigma}=-5.04, V_{sp\sigma}=-5.58$ and $V_{ss\sigma}=-6.77$
in eV\cite{saito}. In the bulk two dimensional case, one can
obtain the intrinsic spin orbit interaction strength $\xi_1\approx
2\xi_0^2\epsilon_s/9V_{sp\sigma}^2\sim 10^{-3}\mathrm{meV}$ at
Dirac points using the above parameters with $\xi_0\approx
4$meV\cite{a7}. We will also include the hamiltonian
$H_{\mathrm{P}}$ to take into account the passivation of the
dangling orbitals at the edges of the graphene nanoribbon. The
specific form of $H_{\mathrm{P}}$ will be given in the section IV.
For the pristine graphene nanoribbons, that is, the non-passivated
ZGNR in which the dangling bonds at the edges are kept intact, we
will apply the open boundary condition. Other extrinsic spin orbit
coupling terms such as the Rashba interaction are not considered
here.

\section{The pristine zigzag graphene nanoribbon}

The effects of the spin orbit coupling on the electronic
properties of the Dirac particles in graphene have been
extensively studied recently\cite{a1,a2,a5,a6,a7,a9}. Since the
energy levels of $\sigma$ bands are well separated from the Dirac
points, one can obtain the low-energy effective hamiltonian
projected to the Hilbert space of the $p_z$ orbital alone by
integrating out the high energy processes involving the $\sigma$
bands\cite{pert}: $H_{\mathrm{eff}}\simeq
H_\pi-H_{SO}H_\sigma^{-1}H_{SO}$. It has been shown that the
nearest neighbor (n.n.) hopping amplitude induced by the SOC is
cancelled by the lattice symmetry and thus the leading
contribution due to the intrinsic SOC results from the effective
next nearest neighbor (n.n.n.) hopping processes of the following
form $i\xi_1 /3\sqrt{3}(\vec{d}_{ik}\times \vec{d}_{kj})c^\dag_i
s^z c_j$\cite{a1,a2}, where $\xi_1$ is on the order of
0.05K\cite{a5,a7}. This effective hamiltonian, so called the KM
hamiltonian, has been generally used to study the edge states of
graphene nanoribbon, which led to the QSHE. However, we want to
point out that since the lattice symmetry is broken at the
graphene edges, one should in principle use the full hamiltonian
instead of the truncated low-energy effective KM hamiltonian. In
doing so, it is generally observed that the characteristic
behavior of the QSHE in the ZGNR depends largely on the edge
geometry and the passivation.

In Fig. 2(a), we have shown the calculated band structure of the
pristine ZGNR with width $N=150$. From the two degenerate
uncoupled compositions of the spin and orbit $\{ p_z\uparrow,
p_x\downarrow, p_y\downarrow, s\downarrow\}$ and $\{
p_z\downarrow, p_x\uparrow, p_y\uparrow, s\uparrow\}$, we have
plotted the former one in Fig. 2(a). Here, we have taken a
relatively large value of the SOC $\xi_0=0.1$eV for the sake of
clarity, since we have noticed that the magnitude of the SOC
hardly affects the qualitative features of the band structures. In
order to investigate the edge states made of the $p_z$ orbital in
detail, we focus on the $\pi$-edge bands within the dashed box of
Fig. 2(a), which are magnified in Fig. 2(b). We have also
calculated the band structure based on the effective KM
hamiltonian, which is shown in Fig. 2(c) and compared to the
result from our model. We have chosen $\xi_1\approx
2\xi_0^2\epsilon_s/9V_{sp\sigma}^2=6.34\times 10^{-4}$eV for the
KM-hamiltonian, which corresponds to $\xi_0=0.1$eV for our
hamiltonian.

By the simple band counting, one can expect to have $(2N-2)\times
4$ bulk bands and the $2\times 4=8$ edge bands composed of the
four atomic orbitals, all of which are two-fold spin degenerate
due to the time reversal and the inversion symmetry. The most part
of the band structure can be understood as a confinement effect on
the bulk two dimensional graphene. The almost flat bands within
the black dashed box and the red bands are the newly introduced
states which are absent in the bulk graphene. They represent the
edge localized states, where the former ones ($\pi$ edge bands)
are mostly made of $p_z$ orbital and the latter ones ($\sigma$
edge bands) mainly consists of $s, p_x, p_y$ orbitals. These
features of the band structures of the ZGNR are consistent with
the previous first principles calculations\cite{cho}. It has also
been widely known that the gapless flat bands of $p_z$ orbital are
formed at $E=0$ within the finite region of $2\pi/3<k<\pi$ in the
absence of the SOC. It is shown in Fig. 2(a) that there exist six
$\sigma$ edge bands with two-fold spin degeneracy and three pairs
of them are almost degenerate so that it seems only three $\sigma$
bands exist. For the stronger SOC, they will split into distinct
six non-degenerate bands. While two pairs of them are well
separated from the $\pi$ edge bands, a pair of the $\sigma$ edge
bands appear quite close to the $\pi$ edge ones. Since two pairs
of edge $\sigma$ bands lie below the $\pi$ edge bands, the Fermi
level of the undoped ZGNR is located below the $\pi$ edge ones as
shown in Fig. 2(a).

At this charge neutral point, we have the edge localized states
coming from the two $\sigma$ edge bands, which become degenerate
at the zone boundary. At a fixed value of $k$ away from the zone
boundary, the two almost degenerate $\sigma$ bands are shown to be
localized at different edges leading to the even number of pairs
of edge states at each edge. Hence the QSHE does not occur. In
contrast to our results, the KM-model exhibits the spin filtered
edge states within the energy window of width $2\xi_1$ around the
band crossing point at the zone boundary and thus one can expect
the QSHE to occur as shown in Fig. 2(c).

The $\pi$-edge bands are shown in Fig. 2(b) and 2(c), where the
red and blue bands consist of the edge states confined to the
right and left edge of the ribbon respectively. The black dashed
bands represent the states, whose amplitudes are spread over both
edges. For another set of the spin-orbit composition $\{
p_z\downarrow, p_x\uparrow, p_y\uparrow, s\uparrow\}$, one can
simply interchange the red and blue colors of the bands. In Fig.
(3), we have shown that the states within the $\pi$-edge bands
($k=2.245, 3.5$) are strongly confined to one of the ribbon edges
with the localization length $\xi\cong a/\ln (-2\cos ka/2)$, which
vanishes as $k$ approaches to the zone boundary ($k=\pi$). In
contrast, the states at $k=2.134$ have a bulk feature, which have
finite amplitudes along the width direction.

By increasing the carrier density, one can adjust the Fermi energy
into the region, where spin filtered chiral edge states made of
the $\pi$ orbitals do exist within the energy window of width
$2\xi_1$ and the QSHE will appear. Here the red and blue bands are
monotonic and move in the opposite directions to each other. With
a further increase of the carrier density, one can raise the Fermi
energy to the band crossing point at the zone boundary. At this
time reversal invariant point denoted by an asterisk in Fig. 2(b),
the edges states are most strongly confined to the ribbon edges
for both the KM model and ours. While the KM model yields the QSHE
near this point with a single pair of edge states at each edge as
shown in Fig. 2(c), our model demonstrates that two chiral spin
bands are non-monotonic and they cross the Fermi level several
times as shown in Fig. 2(b). There exist edge states propagating
in both directions at each edge of the ribbon. Although the number
of bands crossing the Fermi level is odd, one of them(black dashed
one) is always dispersed at both edges manifesting the feature of
quantum confined bulk band. This means that there exist two pairs
of edge states at each edge and hence the QSHE is not feasible in
this energy range.

We have also studied the effect of the spin orbit coupling on the
band structure of the AGNR. In comparison to the ZGNR, the Dirac
cone is located at the $\Gamma$ point ($k=0$), which becomes split
in energy upon the inclusion of the SOC. It is well known that in
the absence of the SOC, the gapless edge bands exist which crosses
$E=0$ at $k=0$ for $N+1$ being an integer multiple of three. Hence
the main difference between the ZGNR and the AGNR lies in the fact
that while there exist a finite range of gapless flat bands for
the ZGNR, there is a single gapless point at $k=0$. In Fig. 4, the
band structure of the AGNR with width $N=152$ is plotted. One can
clearly see that the $\pi$ edge bands are formed around $k=0$
within the energy range of $2\xi_1$ and hence the QSHE is expected
to occur. The amplitudes of the eigenvectors at a fixed value of
$k=0.05$ are plotted as a function of dimer line index for three
different values of $\xi_1=6.34\times 10^{-4}, 1.58\times 10^{-2},
6.34\times 10^{-2}$ in eV. One can notice that in contrast to the
case of ZGNR, the $\pi$ edge states for the AGNR are quite widely
spread. The localization length of the edge states for the AGNR
can be approximately given by $(\gamma_0/\xi_1)a$ with $a$ being
the lattice spacing.

\section{The hydrogen passivated zigzag graphene nanoribbon}

In the section, we will investigate the effect of the hydrogen
passivation on the edge dangling bonds of the ZGNR. Since the
hydrogen atom has a single $s$ orbital, only the $p_x$ and $s$
orbitals of the adjacent carbon atom will have a finite overlap
with the hydrogen atom so that the hamiltonian $H_{\mathrm{P}}$
for the edge passivation can be written as follows \be
H_{\mathrm{P}}&=&\sum_{n}\Big[\tilde{V}_{sp}c^\dag_{p_x,A,N,n}h_{N,n}+\tilde{V}_{ss}c^\dag_{s,A,N,n}h_{N,n}
-\tilde{V}_{sp}c^\dag_{p_x,B,1,n}h_{1,n}+\tilde{V}_{ss}c^\dag_{s,B,1,n}h_{1,n}+\mathrm{h.c.}\Big]\nonumber
\\&&+\sum_{n}\varepsilon_h(h^\dag_{1,n}h_{1,n}+h^\dag_{N,n}h_{N,n}),\ee
where the two hopping and on-site parameters between the carbon
and hydrogen atom are taken to be $\tilde{V}_{sp}=-4.5$,
$\tilde{V}_{ss}=-4.2$ and $\varepsilon_h=-2.7$ in eV\cite{poly}.
The operator $h^\dag_{i,n}$($h_{i,n}$) represents the
creation(annihilation) operator of an electron at hydrogen atom
bonded to the $i$-th carbon dimer line in the $n$-th unit cell.

The band structure of the hydrogen passivated ZGNR is shown in
Fig. 5(a). There exist eight $\sigma$-edge bands(red ones) which
are composed of four pairs of almost doubly degenerate bands. In
comparison to the non-passivated ZGNR, we have an additional pair
of $\sigma$ edge bands, since the hydrogen $s$-orbitals at both
edges have been coupled to the original $s$ and $p_x$ orbitals in
the edge carbon atoms. Since the on-site energy
$\varepsilon_h=-2.7\mathrm{eV}$ of hydrogen atom is quite close to
the band bottom of the $\sigma$ edge band located in the middle
for the non-passivated ZGNR, they strongly interact with each
other and then repel as shown in Fig. 5(a). This makes two
significant effects on the edge state characteristics. Focusing on
the $\pi$-edge bands shown in Fig. 5(b), one can notice that the
general feature of the KM-model is recovered upon hydrogen
passivation. In addition, a newly introduced pair of $\sigma$ edge
bands are placed above the $\pi$ edge bands. This balances the
number of energy bands above and below the $\pi$ edge bands so
that the Fermi level at half filling is placed on the $\pi$ edge
bands.

By comparing the positions of the $\sigma$ edge bands in Fig. 2(a)
and 5(a), one may presume that the $\sigma$-edge bands located
close to $E=0$ mainly affect the energy dispersion of the
$\pi$-edge band.
In order to confirm this scenario, we have studied the effects of
the $\sigma$-edge bands by using the perturbation method, where
the low-energy effective hamiltonian is given by
$H_{\mathrm{eff}}\simeq H_\pi-H_{SO}H_\sigma^{-1}H_{SO}$. The
hamiltonian $H_{\mathrm{eff}}$ can be decomposed into two terms:
$H_{\mathrm{eff}}=H_{\mathrm{bulk}}+H_{\mathrm{edge}}$, where
$H_{\mathrm{bulk}}$ and $H_{\mathrm{edge}}$ for a given $k$ can be
written by
\begin{eqnarray}
H_{\mathrm{bulk}}=H_\pi-\sum_{i\in \mathrm{bulk}}H_{SO}|v_{\sigma
i}\rangle E_{\sigma i}^{-1}(k) \langle v_{\sigma i}|H_{SO}
\nonumber\\H_{\mathrm{edge}}=-\sum_{i\in
\mathrm{edge}}H_{SO}|v_{\sigma i}\rangle E_{\sigma i}^{-1}(k)
\langle v_{\sigma i}|H_{SO}.
\end{eqnarray}
Here $|v_{\sigma i}\rangle$ represents the eigenstate of the
hamiltonian ${\tilde H}_\sigma=H_\sigma+H_{\mathrm {P}}$ in the
$i$-th $\sigma$ band and
$\sum_{i\in\mathrm{bulk}}$($\sum_{i\in\mathrm{edge}}$) stands for
the sum over the bulk(edge) eigenstates. We expect that
$H_{\mathrm{bulk}}$ will reproduce the Kane-Mele hamiltonian and
thus will always produce spin chiral edge bands at the band
center, which is clearly demonstrated in the insets of both Fig.
5(b) and 5(c).

By including the $\sigma$ edge band contribution
$H_{\mathrm{edge}}$ to $H_{\mathrm{bulk}}$, we have obtained the
results denoted by the open circles in Fig. 5(b) and 5(c). The
solid lines represent the exact numerical results of the full
hamiltonian, which give an excellent agreement with those from the
perturbation method. Hence we have clearly demonstrated that the
hydrogen passivation can change the general features of the
$\pi$-edge band by modifying the $\sigma$-edge band profile.

\section{The effective real space hamiltonian for the ZGNR}
In the previous section, we have obtained an effective hamiltonian
$\textrm{H}_\textrm{eff}(k)$ projected to the Hilbert space
spanned by the $p_z$ orbitals in the momentum space using the
perturbaion method. Here we will obtain the real space effective
hamiltonian by applying the inverse Fourier transformation(IFT) to
$\textrm{H}_\textrm{eff}(k)$ and analyze the the spatial
dependence of the on-site energy and hopping parameters. For
instance, if one applies the IFT to the
$\textrm{H}_\textrm{eff}(\vec {k})$ of the two dimensional
graphene including the SOC term, one can obtain the n.n.n. hopping
terms as a leading imaginary hopping process in addition to the
original Dirac hamiltonian leading to the KM-hamiltonian written
by $\textrm{H}_\textrm{KM}=\gamma_0\sum_{\langle ij
\rangle}c^\dag_{i\alpha}c_{j\alpha}+\sum_{\langle\langle
ij\rangle\rangle}i\xi_1/3\sqrt{3}\nu_{ij}s^z_{\alpha\beta}c^\dag_{i\alpha}c_{j\beta}$.

By investigating the real space hamiltonian for the ZGNR, we have
been able to study the effect of the broken translation symmetry
at the ribbon edges and the hydrogen passivation as well. The IFTs
of the $\textrm{H}_\textrm{eff}(k)$ for both the non-passivated
and hydrogen passivated ZGNR have yielded as the leading orders
the spatially dependent on-site potential and the imaginary n.n.n.
hopping amplitude as shown in Fig. 6(a)-(d). We have also checked
the additional n.n. hopping term induced by the SOC, which is
absent in the 2D graphene due to the bulk lattice symmetry. We
note that it is finite but much smaller than that of the n.n.n.
hopping processes at the edges and exponentially decreases away
from the edges approaching to zero which is its asymptotic limit.
Based on the above parameters, we have constructed the following
real space effective hamiltonian to describe the ZGNR \be
\textrm{H}_{\textrm{eff}}=\gamma_0\sum_{\textrm{n.n.}}c^\dag_{\alpha,i,n}c_{\beta,j,m}+i\sum_{\textrm{n.n.n.}}
\xi_1(j)/3\sqrt{3}\nu_{in,jm}c^\dag_{\alpha,i,n}c_{\alpha,j,m}+\sum_{\alpha,i,n}E_0(i)c^\dag_{\alpha,i,n}c_{\alpha,i,m}\ee
where the indices $\alpha(\beta)$, $i(j)$ and $m(n)$ stand for
sublattices, dimer lines and unit cells along the longitudinal
direction respectively. The first term represents the
non-interacting Dirac hamiltonian and $\xi_1(i)$ and $E_0(i)$
stand for the imaginary n.n.n. hopping amplitude and the on-site
potential energy, which depend on the dimer line index. In Fig.
6(a)-(d), we plot $E_0(i)$ and
$\tilde{\xi}_1(i)=\xi_1(i)/3\sqrt{3}$ as a function of the dimer
line index both for non-passivated and hydrogen passivated ZGNR
with a ribbon width $N=20$. We have also performed the similar
calculations for ZGNRs with much large width and obtained
essentially the same curves for the tight binding parameters.

First, we compare the spatial dependence of $E_0(i)$ for the
non-passivated and hydrogen passivated ZGNRs as shown in Fig. 6(a)
and 6(b) respectively. For both cases, $E_0(i)$ has shown a steep
increase as one approaches to one side of the ZGNR for each
sublattice. The rate of increase for the non-passivated ZGNR is
much higher than that of the hydrogen passivated one. We notice
that the bending of the $\pi$ edge bands for the non-passivated
ZGNR originates from the steep confinement potential. Concerning
the $\pi$ edge bands, the edge states near the zone boundary
($k=\pi$) are much more strongly localized than the other states.
Since the on-site energy shows a steep rise at the edges, the
states localized tightly at the edge will be more strongly
influenced and will gain an upward energy shift. This explains the
fact that the edge bands near the zone-boundary are more warped
than the other regions. For the case of the hydrogen passivated
ZGNR, however, the effect of the on-site potential is not
noticeable, since the on-site potential difference between the
edge and the middle of the ribbon is one order of magnitude
smaller than that of the non-passivated ZGNR. In Fig. 6(c) and
6(d), the imaginary n.n.n. hopping amplitudes $\tilde{\xi}_1(i)$
are plotted, which demonstrate the strongly enhanced values near
the edges of both the non-passivated and hydrogen passivated
ZGNRs. For the non-passivated ZGNR, we have found that concerning
the direction of the hopping, the sign of the imaginary n.n.n.
hopping parameter near the ribbon edges is opposite to that inside
the ribbon as shown in Fig. 6(e). For the hydrogen passivated one,
the sign is shown to be identical all over the ribbon just like
the KM hamiltonian as shown in Fig. 6(f). Interestingly we observe
that whether the edge states are localized to one side or the
other can be manipulated by controlling both the sign and the
magnitude of the imaginary n.n.n. hopping parameter near the edge.

\section{Conclusions}
We have studied the effect of the edge termination on the low
energy physics of the ZGNR and AGNR by directly solving the tight
binding hamiltonian which includes all the hopping processes
between $s$, $p_x$, $p_y$ and $p_z$ in addition to the intrinsic
SOC. We have obtained the warped $\pi$-edge bands for the
non-passivated ZGNR and also noticed that the Fermi level lies
below the $\pi$ edge bands, which crosses the $\sigma$ edge bands.
Hence at the charge neutral point, we do not expect the QSHE to
occur. We have shown that by electron doping, one can raise the
Fermi level into the region, where the QSHE can occur.
Interestingly, we have demonstrated that the hydrogen passivation
at the edges of ZGNR can recover the standard features of the
$\pi$ edge bands suggested by the Kane-Mele model. Hence our
observation implies that the ZGNR is a nice example, which
demonstrates the importance of the interplay between the
topological classification based on the bulk property and the edge
geometry.

We have also shown that the warping of the $\pi$ edge bands is due
to the strong influence from the $\sigma$ edge bands located close
to the $\pi$ edge bands, which has been confirmed by the
systematic perturbation analysis. Following the inverse Fourier
transformation, we have been able to obtain the real space
effective hamiltonian. Based on the hamiltonian thus obtained, one
can see that the on-site energy and the effective spin orbit
coupling strength are strongly enhanced as one approaches to the
ribbon edges. The steep rise of the confinement potential leading
to the strong effective lateral electric field can also explain
the warping of the $\pi$-edge bands as well.

\section{Acknowledgments}
This work was supported by the Korea Research Foundation Grant
funded by the Korean Government (MOEHRD, Basic Research Promotion
Fund) through KRF-2008-313-C00243.


\newpage
\begin{figure}
\includegraphics[width=1\columnwidth]{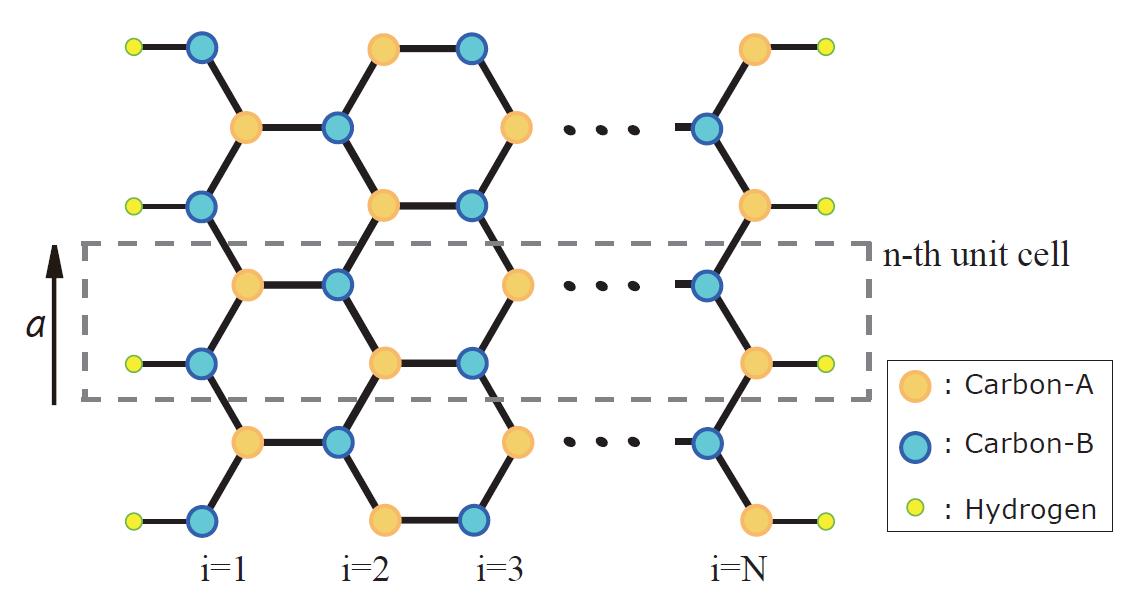}
\caption{(Color online) Hydrogen passivated ZGNR with width N. Two
sublattices of the honeycomb lattice are represented by `carbon-A'
and `carbon B'.}
\end{figure}

\newpage
\begin{figure}
\includegraphics[width=1\columnwidth]{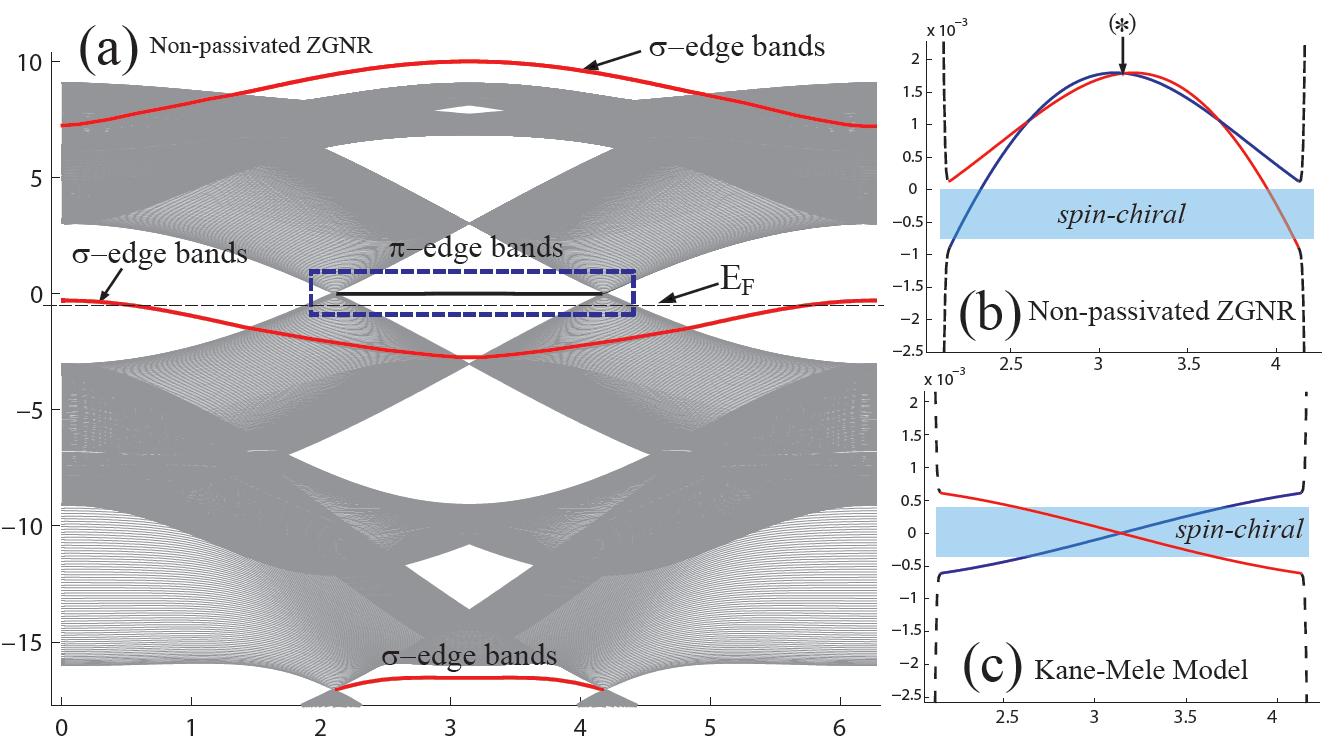}
\caption{(Color online) (a) Band structure of the pristine ZGNR
with width $N=150$ and the atomic SOC strength $\xi_0=0.1
\mathrm{eV}$. Here the $x$-axis represents the dimensionless wave
number $ka$ and the $y$-axis the energy in eV. The red curves
represent the $\sigma$ edge bands made of $s, p_x, p_y$ orbitals.
In the blue dashed box at the band center, we have the $\pi$ edge
bands which look almost flat at this energy scale. (b) The dashed
box in Fig. 2(a) is magnified. (c) The band structure near the
band center obtained by Kane-Mele model with $\xi_1=6.34\times
10^{-4}\mathrm{eV}$. In (b) and (c), the red(blue) bands include
edge states confined to the right(left) side of the ribbon. }
\end{figure}

\newpage
\begin{figure}
\includegraphics[width=1\columnwidth]{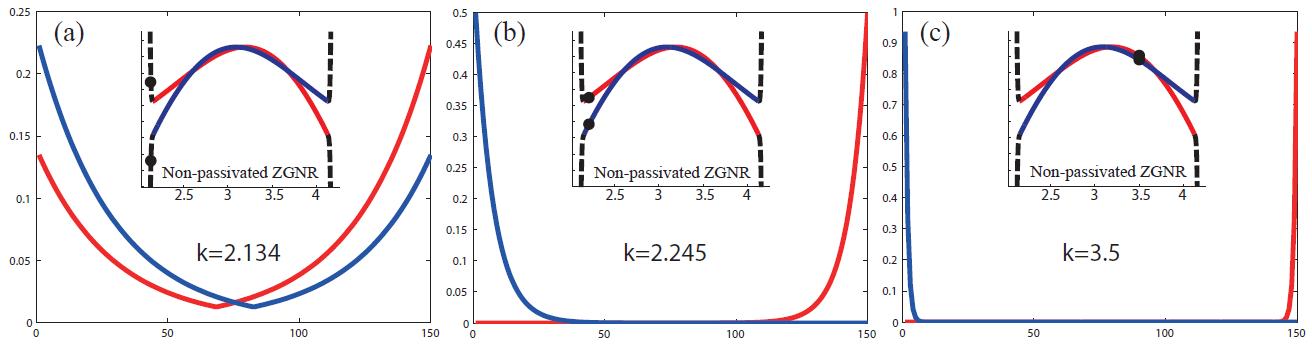}
\caption{Eigenvectors of the $\pi$ edge states at several values
of $k$. The x-axis represents the index of the dimer lines of the
ribbon and the y-axis the absolute value of the amplitude of the
eigenvector at each dimer line. (a) The eigenvector at $k=2.134$,
(b) the eigenvector at $k=2.245$, and (c) the eigenvector at
$k=3.5$. The blue(red) colored eigenvectors correspond to the
blue(red) bands. Here, the blue(red) bands consist of the edge
states localized at the left(right) side.}
\end{figure}

\newpage
\begin{figure}
\includegraphics[width=1\columnwidth]{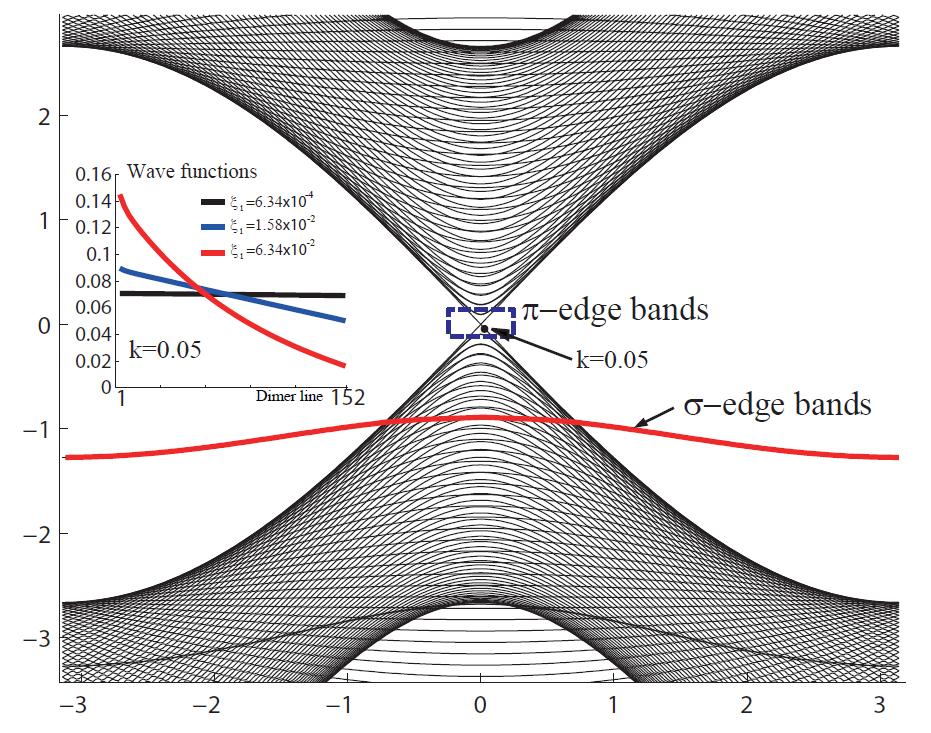}
\caption{(Color online) Band structure of the non-passivated AGNR
with width $N=152$. The red bands represent the $\sigma$ edge
bands which are almost doubly degenerate. Within the dashed box,
we have the edge localized $\pi$ bands. In the inset, the
amplitudes for the eigenvectors at $k=0.05$ are plotted as a
function of dimer line index for three different values of
$\xi_1=6.34\times 10^{-4}, 1.58\times 10^{-2}, 6.34\times 10^{-2}$
in eV. One can notice that the edge states are quite broadly
dispersed rather than localized to the edge.}
\end{figure}

\newpage
\begin{figure}
\includegraphics[width=1\columnwidth]{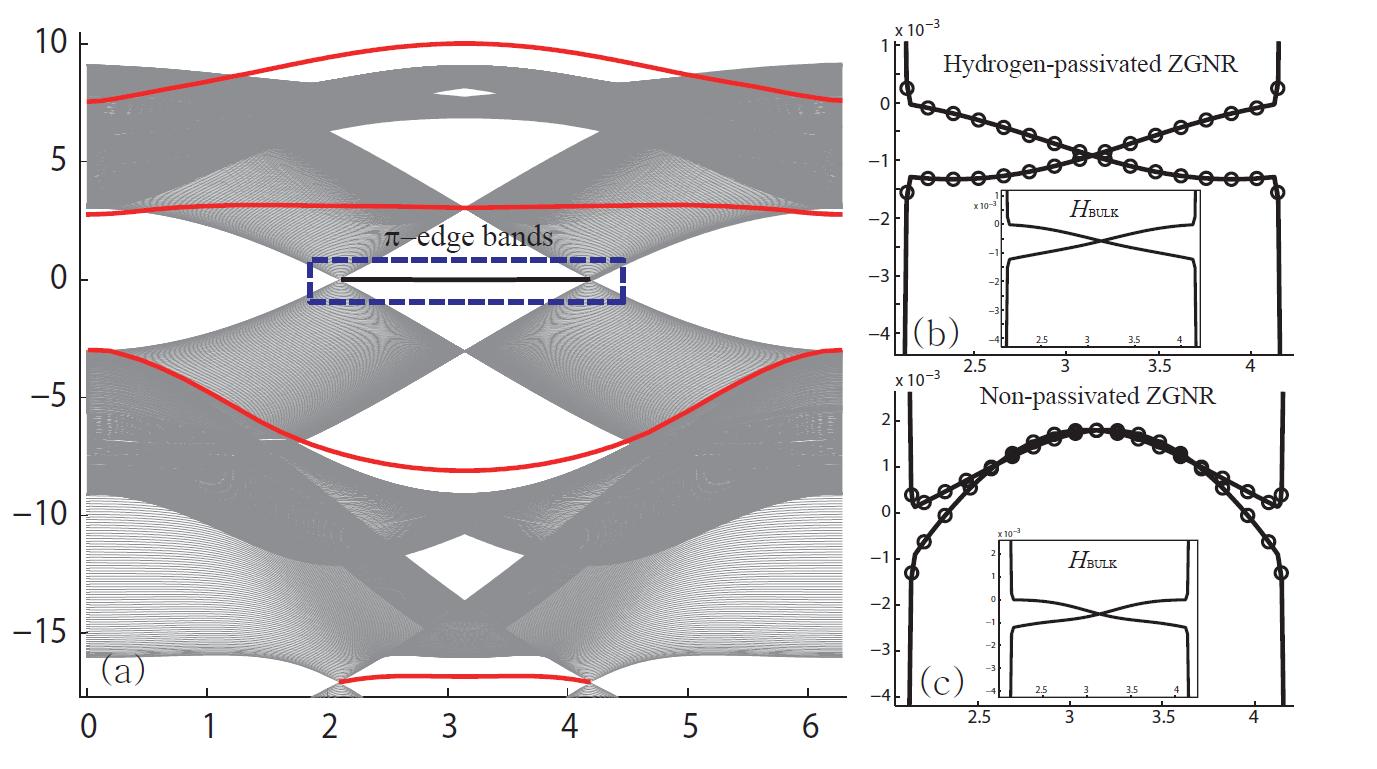}
\caption{(Color online) Band structure of the ZGNR with width
$N=150$ (a) with hydrogen passivation. The passivation parameters
are taken to be $\tilde{V}_{sp}=-4.5$, $\tilde{V}_{ss}=-4.2$ and
$\varepsilon_h=-2.7$ in eV. The red bands represent the
$\sigma$-edge states. (b) The solid lines represent the magnified
view of the $\pi$ edge bands in the blue dashed box. The open
circles correspond to the result obtained from the perturbation
method. (c) The solid lines represent the $\pi$ edge bands without
passivation. The open circles correspond to the result obtained
from the perturbation method for the non-passivated ZGNR. The
insets of (b), (c) stand for the results obtained from
$H_{\mathrm{bulk}}$.}
\end{figure}

\newpage
\begin{figure}
\includegraphics[width=1\columnwidth]{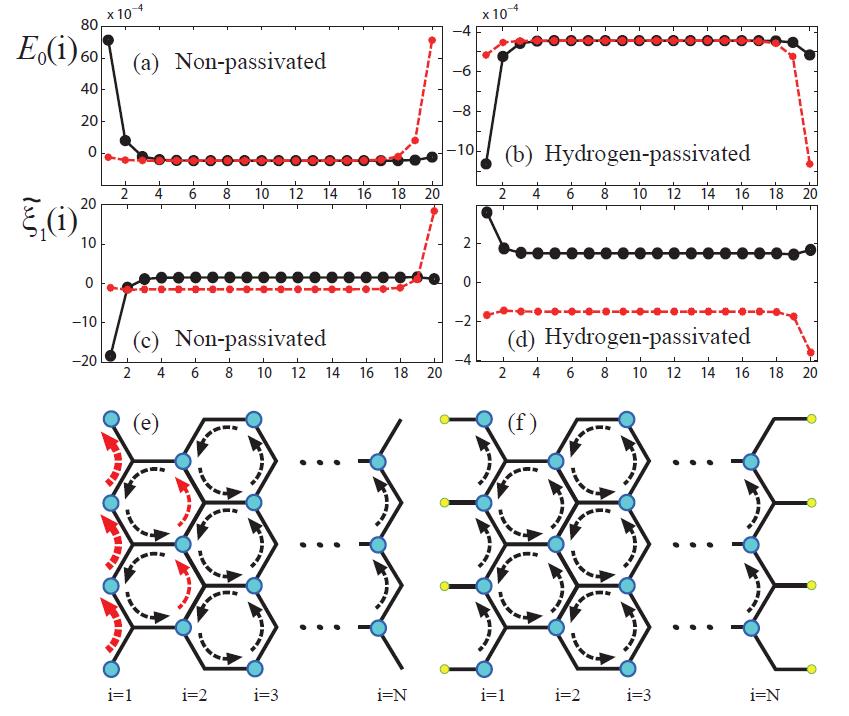}
\caption{(Color online) The spatial dependence of the on-site
energy $E_0(i)$ and the effective n.n.n hopping parameter
$\tilde{\xi}_1(i)$ in eV for the non-passivated and hydrogen
passivated ZGNR with width $N=20$ and
$\tilde{\xi}_1(i)=\xi_1(i)/3\sqrt{3}$. $E_0(i)$ and
$\tilde{\xi}_1(i)$ are plotted in (a) and (c) for the
non-passivated ZGNR, while in (b) and (d) for the hydrogn
passivated one. The dashed red and solid black curves are plotted
for the A and B sublattices respectively. In (e) and (f), we
depict schematically the signs and magnitudes of
$\tilde{\xi}_1(i)$ for the counterclockwise hopping processes. The
black(red) arrows mean that sign of $\tilde{\xi}_1(i)$ is $+(-)$
and their thickness represents the magnitude of
$\tilde{\xi}_1(i)$. }
\end{figure}
\end{document}